\documentclass[11pt]{article}

\usepackage{bm}
\usepackage[normalem]{ulem}
\usepackage{amsmath,amsbsy,amsfonts,amssymb,graphics,epsfig,float,mathrsfs,amsthm,braket}
\usepackage{caption}
\usepackage{subcaption}
\usepackage{comment}
\usepackage{authblk}
\usepackage[capitalize]{cleveref}
\usepackage[normalem]{ulem}

\newcommand{\s}[1]{{\textsf{\textbf{#1}}}}

\begin{document}

%%%% Article title to be placed here
\title{\s{Design of an engaging-disengaging compliant mechanism by 
using bistable arches}}
\author{ \textsf{Mehul Srivastava\footnote{All these authors contributed equally} $^\dagger$, Trishna Gunna$^{*\dagger}$, Makarand Kandiyaped Serkad$^{*\dagger}$, Manu Sebastian$^\dagger$, and Safvan Palathingal $^\dagger$}}
\date{
{\it $^{\dagger}$Department of Mechanical and Aerospace Engineering,\\ Indian Institute of Technology Hyderabad, Telangana, India}\\[2ex]
% {\it * These authors contributed equally }\\[2ex]
%\today
}
%\author[1]{\textsf{Safvan Palathingal}}
%\author[2]{\textsf{Dominic Vella}}

 \maketitle
\hrule\vskip 6pt

%%%% Abstract text to be placed here %%%%%%%%%%%%
\begin{abstract}
Compliant mechanisms utilise  elastic deformation of their segments to transmit motion or force. The utility and behaviour of specific compliant mechanisms can be enhanced by introducing an engaging and disengaging ability with its elastic segments. Towards this, we present an engaging-disengaging compliant mechanism (EDCM) that can switch its stiffness between infinite and zero. The design of the EDCM is based on bistable arches and a locking mechanism. We describe its working, identify its design parameters, and use analytical expressions to arrive at its dimension. The design is verified by detailed finite element analysis and experiments on a 3D-printed prototype. Three alternate designs that lead us to the final mechanism are also briefly discussed.
\end{abstract}

\vskip 6pt
\hrule
\vskip 6pt
%\maketitle

Locking and unlocking mechanisms find extensive use across diverse domains of robotics \cite{Plooij2015}, where they enable controllable engagement and disengagement of joints such as robotic knees in legged systems, as well as in microelectromechanical systems (MEMS) \cite{Xu2016}, where they serve to isolate structural components or dissipate shocks under high-vibration environments.
% Locking and unlocking mechanisms have a wide range of applications in the field of robotics \cite{Plooij2015}, for locking and unlocking of joints like knees in walkable robots, and MEMS devices \cite{Xu2016}, for isolating parts or absorbing shocks in high vibration environment. 
% \saf{It would be nice to address why there are these applications rather than just saying there are; please add a sentence or two}. 
%ok
% These mechanisms can be classified as mechanical\cite{10.1115/1.4033037cite1, 8632705-mechanical-2cite2,5980276cite4}, frictional \cite{Aghili2006} and compliant \cite{ZHAO2022105083-compliantcite5} types based on the design of engagement and disengagement. 
The type of engagement and disengagement in these mechanisms are broadly based on rigid linkages \cite{Lu2016, Chung2019, Van2011}, frictional elements \cite{Aghili2006}, and compliant segments \cite{Yin2022}. In this work, a compliant contact-aided locking and unlocking mechanism is conceptualised, i.e., a mechanism that can switch its stiffness between infinite and zero. The stiffness is toggled by flipping bistable arches that are part of the mechanism between their stable positions. Hereafter, this Engaging-Disengaging Complaint Mechanism is referred to as EDCM for simplicity. The primary motivation is to introduce the EDCM between two points on a segment of any existing compliant mechanisms. This addition allows the segment to be either compliant or rigid based on the state of the EDCM, which could change the behaviour of the parent compliant mechanism.

% \emph{Might remove this: In this work, we conceptualise  a contact-aided compliant locking and unlocking mechanism which can be used to engage or disengage a member from the rest of the mechanism by means of flipping internal bistable arches between their two stable positions. Hereafter, we call it as Engaging Disengaging Complaint Mechanisms (EDCM). %This mechanism changes the behavior of a system by changing the stiffness between two selected points, from rigid (as if a rigid link connects the two points) to free (the two points have no connection to each other) and vice versa. 
% This mechanism changes the behavior of a system of two points, from rigid (as if a rigid link connects the two points, hence transfer deformation) to free (the two points have no connection between each other, hence retain deformation) and vice versa, with the aid of contact via bistable arches.}

\begin{figure}[!htbp] 
    \centering
\includegraphics[width=\textwidth]
    {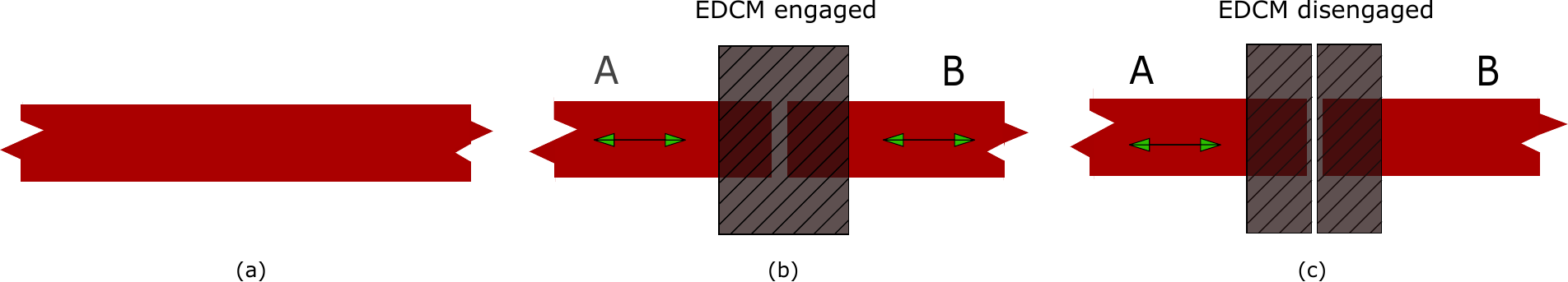}
    \caption{EDCM engages and disengages to (b) transfer and (c) not transfer motion between points A and B. }
    \label{fig_introduction}
\end{figure}

In other words, the primary role of the EDCM is to transfer or retain deformation. \Cref{fig_introduction} 
explain the intended function of the EDCM. In its engaged state, the whole mechanism behaves as a rigid entity. This implies no relative displacement between A and B along the line AB (See \cref{fig_introduction}b). Displacing A or B by a certain amount along the joining direction leads to the movement of the other point in the same direction, and by the same amount. In the disengaged state, the mechanism is virtually non-existent. The points A and B are free to move with respect to each other. Displacing A does nothing to the position of B, and thus, the deformation is retained.

The EDCM should also have a short switching time, ease of control, and compactness. These requirements are met with bistable arches which switch between two stable states. However, having compliant bistable arches in the EDCM, it becomes difficult to completely transfer the deformations between the two points in their engaged state due to the deformation in the arches. Hence, a locking mechanism is introduced within the EDCM to address this problem. Furthermore, the mechanism should also be able to disengage completely to prevent deformations from being transmitted in its disengaged state, which can be achieved relatively easily by bistable arches.

In  \cref{sec2}, the key design element of EDCM is introduced. Further, three preliminary designs are discussed along with their limitations, which led to a final design. The final design is described and its critical design parameters are identified. In \cref{sec3}, the bistable arches involved in the design of EDCM are designed. In \cref{sec4}, the design is validated by finite element analysis(FEA) and a prototype is built.

% We plan to create an Engaging-Disengaging Compliant Mechanism (EDCM). This mechanism will be used to transfer or retain deformation. In its engaged state, the whole mechanism behaves as a rigid entity. This implies no relative displacement between P and Q (See figure \ref{intro:fig}b). Displacing P or Q by a certain amount along the joining direction leads to the movement of the other point in the same direction, and by the same amount. In the disengaged state, the mechanism is virtually non-existent. The points P and Q are free to move with respect to each other. Displacing P does nothing to the position of Q and thus the deformation is retained.
\par

\section{Design principle}\label{sec2}

\begin{figure}[!htbp] 
    \centering
    \includegraphics[width=\linewidth]{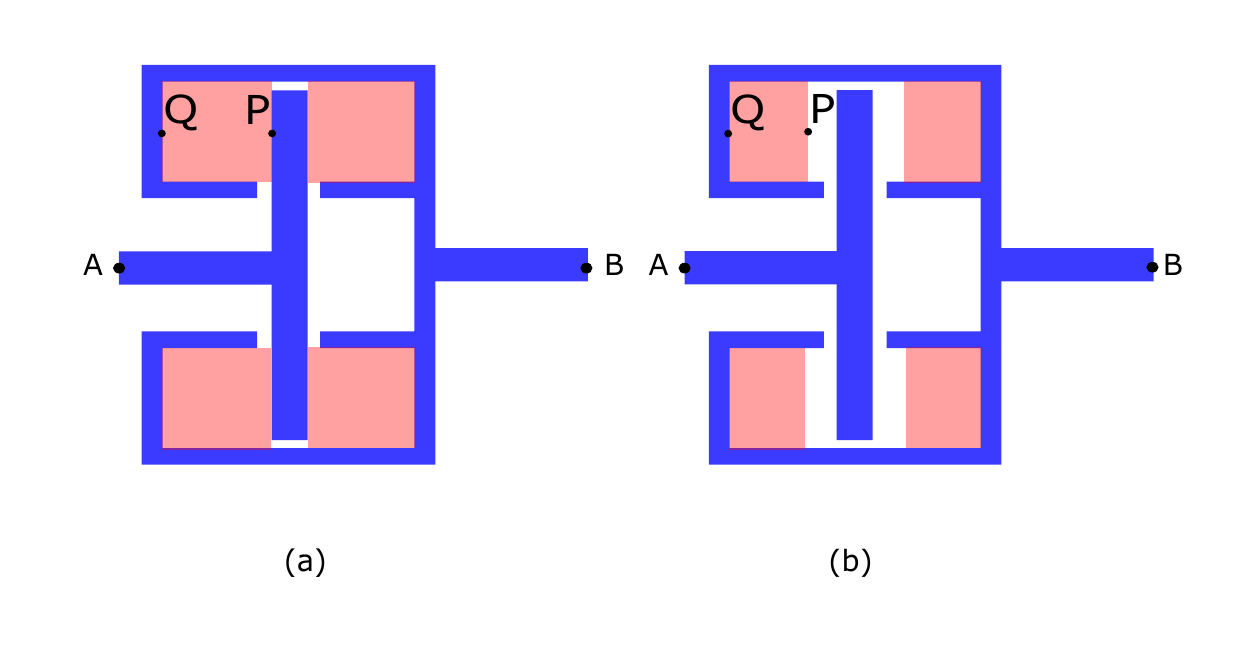}
    \caption{EDCM with (a) EDCMS in locked state (b)EDCMS in free state }
    \label{fig_EDCM}
\end{figure}

\Cref{fig_EDCM} shows how the EDCM can be included in an elastic segment. The switching element inside the EDCM (EDCMS), is the key to the functionality of EDCM. Four EDCMS designs are presented in this section. While all four designs are functional, the stiffness in the locked state for the fourth design is significantly higher than the first three designs.
% This section explains the different design attempts we made for the EDCMS, and the drawbacks of each, which eventually led to the final design.

\begin{comment}
Explaining how the EDCM works
    Upon switching the central arches of the 4 EDCMS, there is space created inside the EDCM, which allows free movement of one point with respect to the other. When the EDCMS is locked (side arches have switched) the EDCM is in its rigid state and deformation is transferred between the two points.
\end{comment}

\subsection{Preliminary designs of EDCMS}
% We initially devised mechanisms in an effort to produce the desired result. These mechanisms, however, did not have perfect locking. They were deforming even when they were intended to be locked and rigid. 
As mentioned, the first three designs in their locked state do not have a considerable stiffness value. In other words, they have some compliance in the engaged state of EDCMS.
The final design has minimal compliance and perfectly switches between the rigid and free states. 

\begin{figure}[!htbp] 
    \centering
    \includegraphics[width=\textwidth]{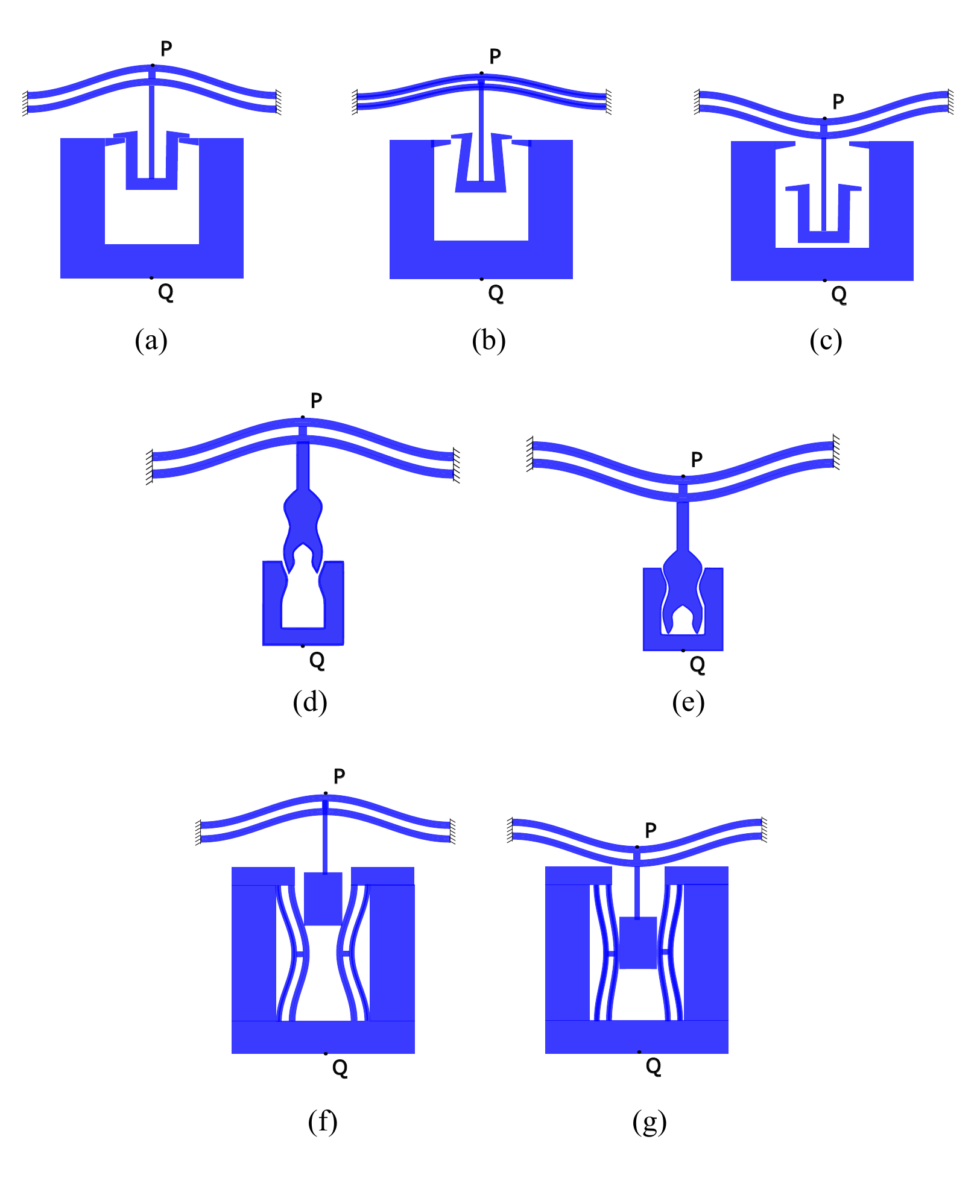}
    \caption{Preliminary designs: Fig. 3 a-c show the first design which is able to resist deformation only for small loads. Fig. 3 d-e show the second design, which reduces the resistance for the clip return but has imperfect locking due to its geometry. Fig. 3 f-g show the third design, in which the resistance for return was further reduced.}
    \label{fig_merged_mechs}
\end{figure}

All the EDCMS presented here have at least a bistable arch and a holder. P is considered as the point at the arch's midspan and Q is at the holder's midpoint, such that the line segment PQ is vertical as shown in \cref{fig_merged_mechs}. P is desired to be stationary when a load is applied in the PQ direction and when the mechanism is locked. Thus, P and Q behave like points on a rigid link (in the direction PQ). When the mechanism is unlocked, i.e., when the central arch is switched downward (\cref{fig_merged_mechs}c), the Point P moves down from its initial position. This corresponds to a free EDCM (see \cref{fig_EDCM}).
% Note that while analyzing the following mechanisms, we apply a load whose magnitude varies linearly with time starting from zero \blue{in the PQ direction}.

%In (figure 2a), the point P is not allowed to move by the EDCMSs, we term this state of the EDCMS as "locked". In the (fig 2b), the central arches of the EDCMSs are flipped. This gives P freedom to move around in the confined space. In this case, the mechanism is "unlocked". When the mechanism is locked, P must not move when load is applied in the PQ direction. Therefore, P and Q behave like points on a rigid link. When the mechanism is unlocked, P is allowed to switch between it's two stable states

% \subsubsection{Mechanism 1}
A latch mechanism is included in the first design to avoid arch deformation during the engaged state as shown in \cref{fig_merged_mechs} a-c. 
 The load acts at P, and the mechanism resists deformation until the load reaches a threshold value required to overcome the latch. For a load of magnitude larger than the threshold value, the arch snaps through to its second state given in \cref{fig_merged_mechs} c.
It is to be noted that this mechanism can hold its rigidity only for small values and cannot be assumed to be perfectly rigid. Furthermore, it is not exactly a switch since it cannot flip back once the mechanism flips or the force applied is greater than the threshold force. So this design could only be used if it needs to be switched only once and the magnitudes of the forces involved are smaller than the threshold value required to snap through the latch. 

% \textcolor{red}{(or large?)}.

The second design depicted in  \cref{fig_merged_mechs} d-e reduces the resistance
% \textcolor{red}{(grammar error)} 
while returning to its as-fabricated state but improvises on the resistance to deformation when locked.
% \textcolor{red}{(Compared to first design this one has less resistance to deformation when locked. So, should we change this line?)}
The latch is replaced with a clip that resembles an hourglass. Due to its shape, the clip resists deformation when locked and facilitates switching when the mechanism moves from unlocked to the locked state. However, the curved contact surfaces cause easy slipping, so the mechanism was not perfectly rigid in the locked state.

The third mechanism's central characteristic is the addition of two snapback arches perpendicular to the existing switching arch as shown in \cref{fig_merged_mechs} f-g. This mechanism was designed to make the switch-back of the central arch easier. 
The block-like bottom portion extended from the central arch pushes the side snapback arches away as it comes down during the switch. The side arches make switching from locked to unlocked state challenging but easier to switch back. However, the problem of small deformations persists and it is not as rigid as the design presented next.

\subsection{EDCMS}
\begin{figure}[!htbp] 
    \centering
    \includegraphics[width=0.6\textwidth]{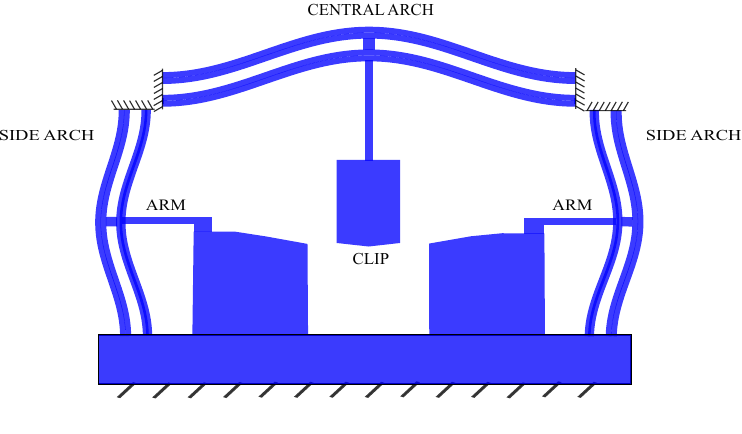}
    \caption{The labelled schematic of the final mechanism.}
    \label{fig_finalmechlabel}
\end{figure}

The final EDCMS design shown in \cref{fig_finalmechlabel} is inspired by the last design presented in the previous section. The side snap-back arches were replaced with bistable arches and two arms were attached at their midspan. These two arms prevent the clip and central arch from moving down in their locked state.

\begin{figure}[!htbp] 
    \centering
    \includegraphics[width=\textwidth]{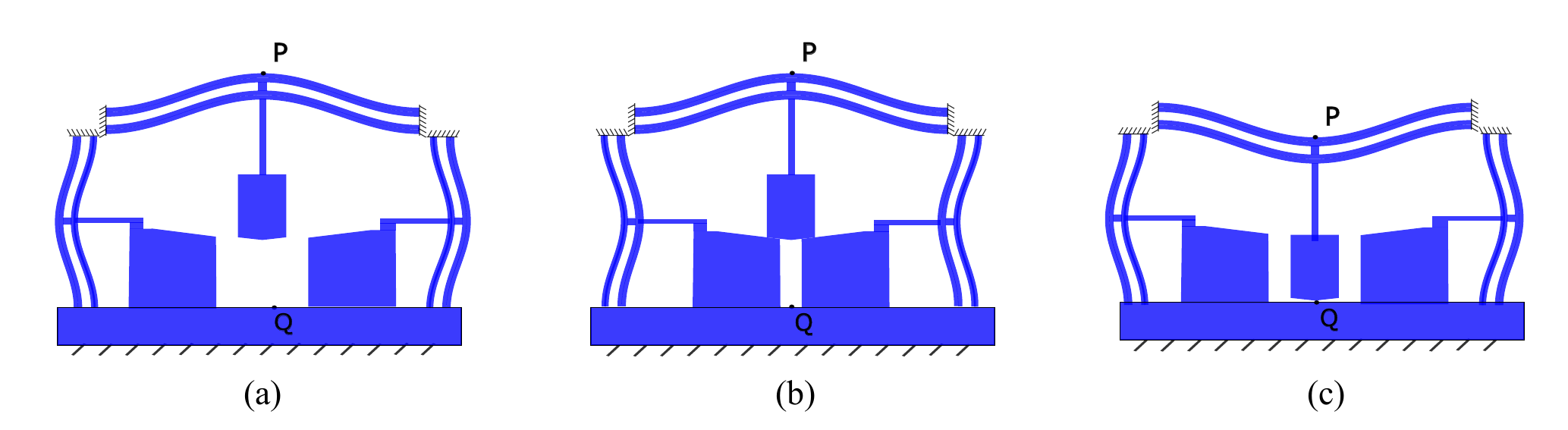}
    \caption{The stable states of Final design. (a) The first stable state of the mechanism when the lock is switched off, (b) The first stable state of the mechanism in which any deformation is resisted as the lock is switched on, and (c) The second stable state.}
    \label{fig_mechfinal}
\end{figure}

When the mechanism is in its as-fabricated state (\cref{fig_mechfinal}a), applying a load at the midspan of the central arch downwards results in free switching (\cref{fig_mechfinal}c) to the unlocked state. However, when the side arches are flipped towards the centre (\cref{fig_mechfinal}b), the two arms hold the clip attached to the central arch and any deformation through the midspan of the central arch is prevented, thereby locking the mechanism. 

It is to be noted that the state of the mechanism is controlled by toggling the two side arches. The central arch can easily switch to its unlocked state and back when the side arches are flipped away from the centre, unlike the previous designs with high resistance while switching back. Furthermore, this design achieves substantial rigidity in the locked state.

To ensure rigidity when the mechanism is locked, there should be no free space between the ends of the clip and the arms. The central arch is designed carefully to satisfy this requirement. The bottom edge of the clip has a slight incline to alleviate unnecessary friction when the clip and the arms engage.

\begin{figure}[!htbp] 
    \centering
    \includegraphics[width=0.7\textwidth]{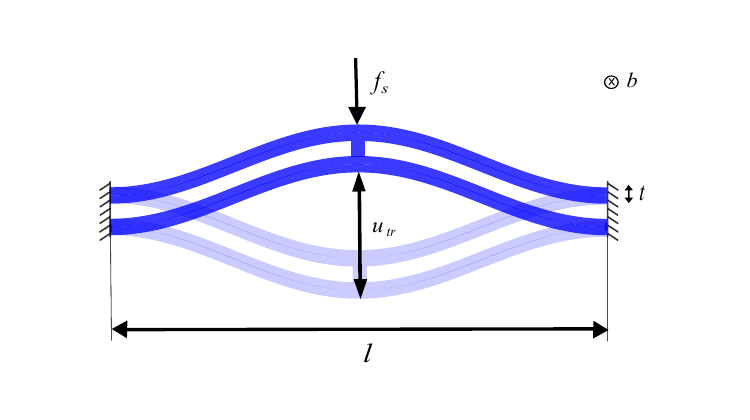}
    \caption{The design parameters considered for the bistable arch. \(f_s\) denotes the minimum switching force, \(u_{tr}\) denotes the travel whereas l, b and t denote the span, width and the depth of the arch respectively. }
    \label{fig_designparams}
\end{figure}

Designing two side and the central bistable arches, which is well studied
in the literature\cite{Palathingal2019,Palathingal2018}, is vital to realising this design. Consider the bistable shown in \cref{fig_designparams}
with a span, \emph{l}, the mid-rise, \emph{$h_{mid}$}, the second moment of area of the cross-section, \emph{I}. Travel, the distance the midpoint of the arch moves between the two stable states, $u_{tr}$, and the switching force, \emph{$f_{s}$}, required to switch from the locked to unlocked state are given by\cite{Palathingal2017}:

\begin{align}\label{eq_fs}
    \frac{f_sl^3}{EIh_{mid}} &= 1486.57,\\
    \label{eq_utr}
    \frac{u_{tr}}{h_{mid}} &= 1.98.
\end{align}
 \cref{eq_utr,eq_fs} are utilized to arrive at the dimensions of the EDCMS in the next section.

%A new schematic that shows the working of the mechanism. This should look similar to the final 3-prototype figure that will include in the Simulation and Prototype section. By using this figure explain the working of the mechanism. 

%It might be also worthwhile to include one of the earlier deisgns to highlight what the current design does very well.

%Identify the key parameters of the design.

\section{Design of bistable arches}\label{sec3}
%This section explains how to design bistable arches. Expressions from the literature etc.

\subsection{Central Arch}\label{subSec2}
The central arch is designed first. It is intended to calculate the dimensions of the bistable arch for a switching force of $17$ N and a travel of $16$ mm. The depth of the arch, \(t\) was taken to be $1$ mm to enable easy 3D-printing. By using \cref{eq_utr}, the following is obtained
\begin{equation}\label{eq_central_utr}
    h_{mid} = \frac{u_{tr}}{1.98} = 8.08~ \text{mm}.
\end{equation}
By substituting these values and by taking \(E\) to be $2.4$ GPa  in \cref{eq_utr}, the following is obtained
\begin{equation}\label{eq_central_fs}
    \frac{l^3}{b}=\frac{1486.57Eh_{mid}t^3}{12}=0.141
\end{equation}
The values of \(l\) and \(b\) are taken to be $70$ mm and $2.5$ mm, respectively  such that the $\frac{l^3}{b}$ ratio is satisfied approximately ($\frac{l^3}{b} = 0.137$). This need not be accurately taken if it is constraint on maintaining the switching force to be exactly $17 N$.

\subsection{Side Arches}\label{subsec2}
%\textcolor{blue}{It'll be the same as the main arch, but there has to be a relation between half the span length of the main arch and the Utr of the side arch, right$?$} 

Following an approach similar to the central arch, the design parameters of the side arches in this design are obtained. It is to be noted that the dimensions for both the side arches are the same. 

The dimensions for both the side arches are calculated for a switching force of $70$N  and a travel of $8mm$. The travel is chosen in a way that when both the side arches are flipped inwards, the two arms come close and directly under the clip attached to the central bistable arch, thereby restricting any possible deformations. The depth of the arch, \(t\) is taken to be $1$ mm, and \(E\) as $2.4$ GPa. Thus,
\begin{equation}\label{eq_side_utr}
    h_{mid} = \frac{u_{tr}}{1.98} = 4.04~ \text{mm}
\end{equation} Then by using \cref{eq_fs},
\begin{equation}\label{eq_side_fs}
    \frac{l^3}{b}=\frac{1486.57Eh_{mid}t^3}{12}=0.0166.
\end{equation}
As before, span and the width are approximately chosen as $35mm$ and $2.5mm$ respectively ($\frac{l^3}{b} = 0.0171$). 

The design is verified with a central arch of span \(70 \) mm, mid-rise  \(8\) mm, depth  \(1\) mm and width  \(2.5\) mm, and the  side arches of span  \(35 \) mm,  mid-rise of \(4\) mm, depth of \(1\) mm and width of \(2.5\) mm with simulations and a prototype in the next section.

\section{Simulation and Prototype}\label{sec4}

%Include the FEA of the design for a given dimesion. Also photographs of the working prototype. 
In order to check the working of the mechanism, finite element analysis (FEA) of the design was carried out in ABAQUS. A quasi-static analysis of the mechanism was performed with a geometric non-linearity model (NLGeom-ON). 

For the FEA, the material property was taken that of Onyx, a polymer used for 3D-printing. Onyx has a density of \(1200kg/m^3\), Young's Modulus of \(2.4GPa\), and Poisson's ratio of \(0.3\). The section is modelled as solid, homogeneous, elastic, and isotropic. 

% To test the final design in both the cases, different boundary conditions and loads were applied at various points on the mechanism. And the variation of either force or displacement at concerned points was plotted against time depending upon whether the EDCM is engaged or disengaged.

% %the reference "concerned points" for P and Q can be changed

% \subsection{EDCM Engaged}

\begin{figure}[!htbp]
    \centering
    \includegraphics[width=\textwidth]{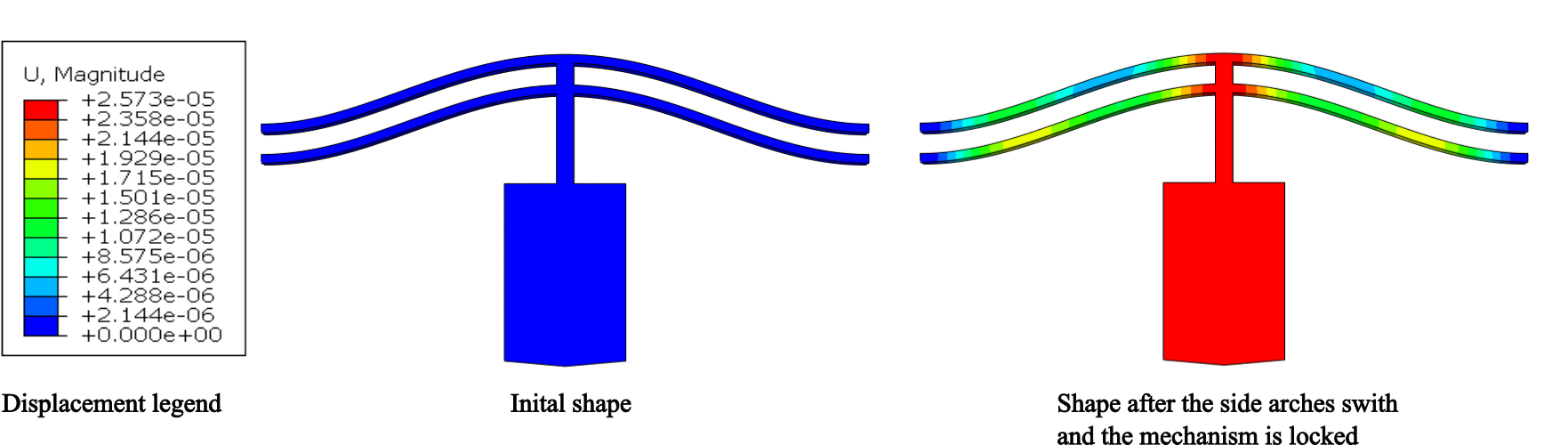} 
    \captionsetup{justification=centering}
    \caption{Displacement of the central arch as the side arches switch}
    \label{fig_feaside}
\end{figure}

In the locked case, it is first analyzed how switching the side arches and locking the mechanism affects the position of point P. Since Q is connected to the holder which includes the side arches, Q will remain unaffected by switching of the side arches. 
% Then we analyse how position of P varies in the locked case when a load is applied on it.
% \subsubsection{Side Arches Switch to Lock the mechanism}
\Cref{fig_feaside}  shows (side arches not shown) that as the side arches switch and lock the mechanism, the central arch as well as the clip move by an amount of $2.5 \times 10^{-5} \,m$. This is $1.5 \times 10^{-4} \,\% $ of  the travel of the central arch and hence can be safely neglected.

%The percentage part I am introducing

%Include a graph showing variation in the positions of P and Q with time in the same plot. Plot Displacement v/s Time. One curve for P and one for Q. Plot 8c (from t=0s to t=1s shows d-t for point P)

\begin{figure}[!htbp]
    \centering 
    \includegraphics[width=0.7\textwidth]{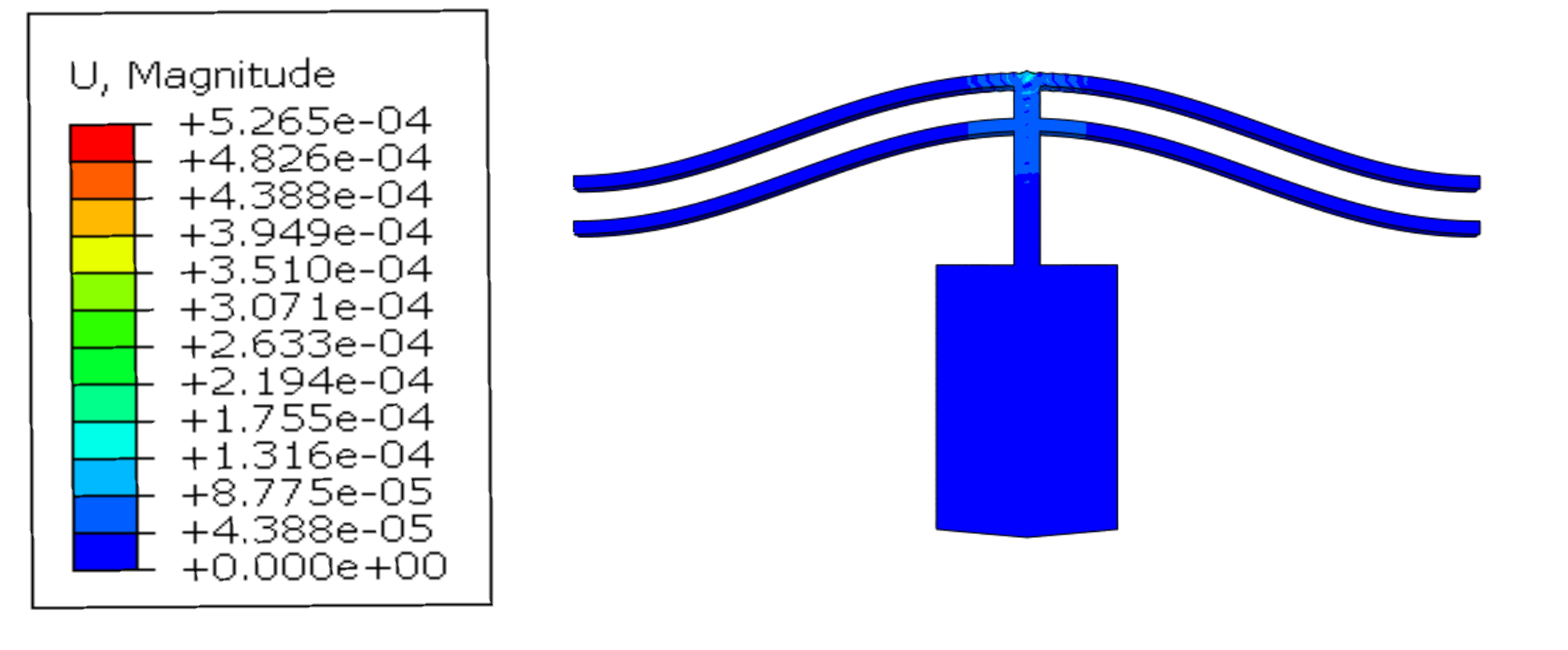} 
    \caption{Displacement of various points of the central arch after mechanism is locked and $50$ N force is applied at the midpoint of its midspan. }
    \label{fig_feacentral}
\end{figure}

\begin{figure}[!htbp] 
    \centering\includegraphics[width=0.9\textwidth]{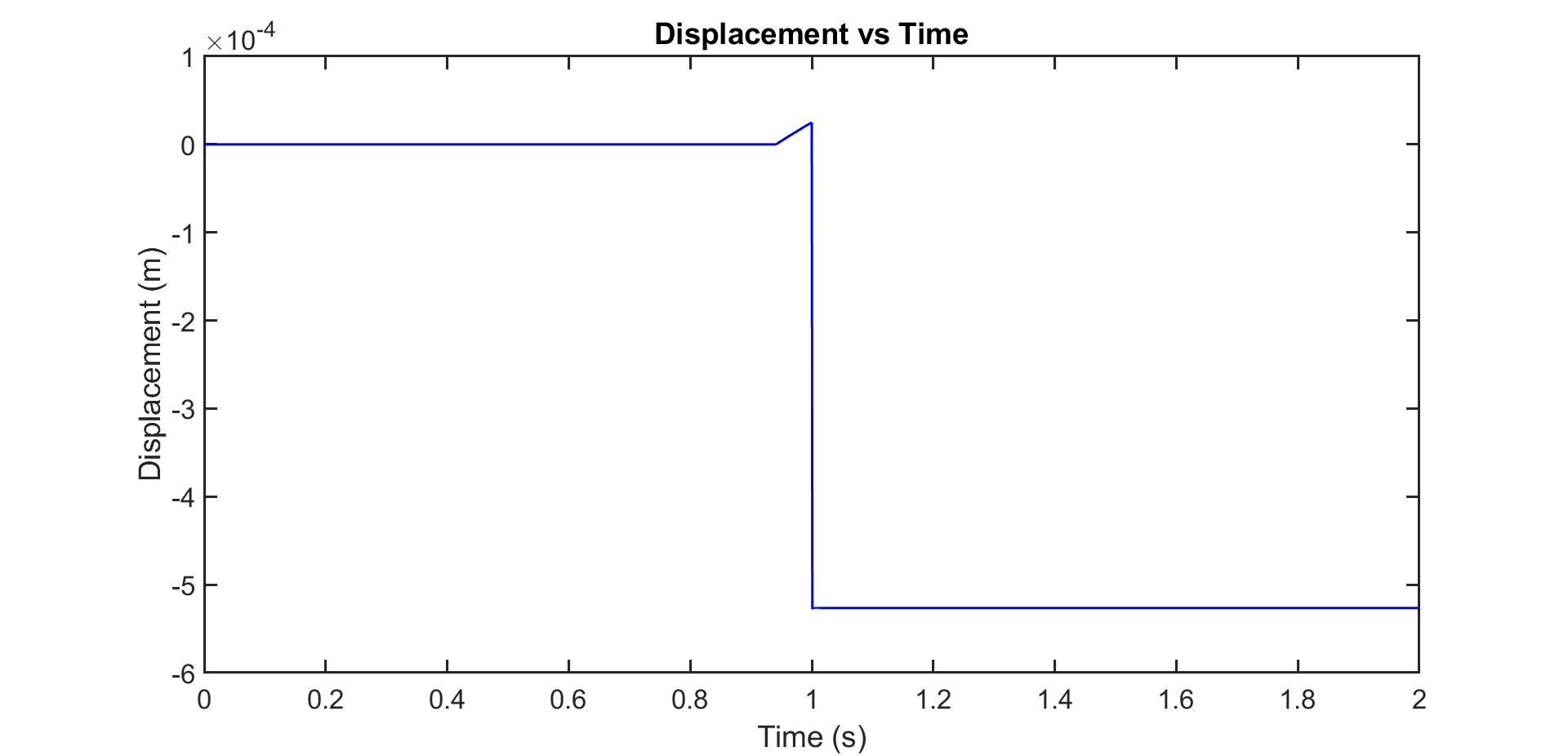}
    \caption{Displacement of the midpoint of the midspan plotted against time as the side arches switch to lock the mechanism and a linearly varying load from zero to 50N load is applied. From time t=0s to t=1s the side arches switch. From t=1s to t=2s the load is applied.}
    \label{fig_plotdisp}
\end{figure}

% \subsubsection{Load is applied at P to test transfer of deformation}
Next, the variation in the position of P when in the locked case is analyzed when a load is applied on it.
To test the rigidity of the mechanism, a load of $50N$ is applied at P while the mechanism is locked. It can be observed that this force is significantly larger than the switching force of the arch.
% Here, we use the logic that if the entire mechanism behaves rigidly, i.e., if it behaves as if the point P and Q are connected by a rigid link, then the deformation will be transferred. Following this, we fix the position of Q and apply load at P directed from P to Q along the line PQ. 
It is observed in \cref{fig_feacentral} that P does not deform. 
%"small percentage of the travel of the central arch" is something I made up. Not sure if it is correct
% The following image shows the variation in the displacement of the central arch after $50N$ load is applied in the locked state of the mechanism.
The small deformation at the point of application of the load is expected. This is clearly visible in \cref{fig_plotdisp} that has the  displacement at point P as a function time. 
From time 0 to 1, the side arches engage and from 1 to 2 the arch has a load of $50N$ at P.
% We find that the the central arch does not displace, however, \\
%We again plot the variation in the positions of P and Q in the same graph and notice relative movement. We find that although Q does not move at all, P shows a displacement of $xxxx\:mm$ which is $yy.yy \%$ of the travel of the central arch and hence may be neglected
% We plot the displacement of the midpoint of the midspan v/s time as the side arches switch and 50N load is applied.
The midpoint of the midspan moves down by an amount of $5.2 \times 10^{-4} \,$m and then stays there even as the applied load reaches a magnitude of $50$ N.

\begin{figure}[!htbp]
\centering
\includegraphics[width=0.8\textwidth]{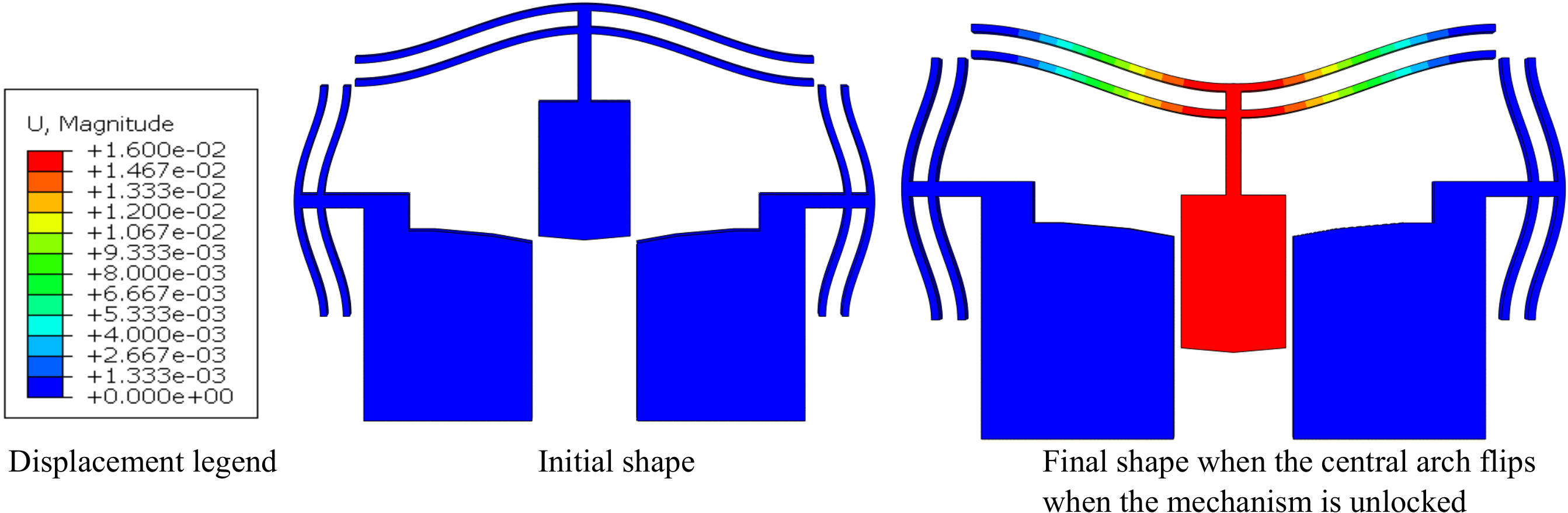} 
\caption{Displacement of the arch when the mechanism is in the unlocked position.
}
\label{fig_feamech}
\end{figure}

To complete the analysis, the deformation is analyzed when the EDCM disengages. For this, P is moved downwards by an amount equal to the travel of the central arch, thereby switching it as shown in \cref{fig_feamech}. The switching force and travel obtained is as per our design.

% We plot the variation of force applied at the midpoint of the midspan v/s time as the arch switches in Fig. \ref{fig_plotforce} and find that there are no discrepancies
% \begin{figure}[!htbp]
% \includegraphics[width=0.9\textwidth]{unlocked_flipping_graph.png} 
% \caption{Graph of force variation at the midpoint of midspan of the central arch as it switches inside the EDCM}
% \label{fig_plotforce}
% \end{figure}
\subsection{Prototype}
\begin{figure}[!htbp]
\begin{subfigure}{0.32\textwidth}
\includegraphics[width=\textwidth]{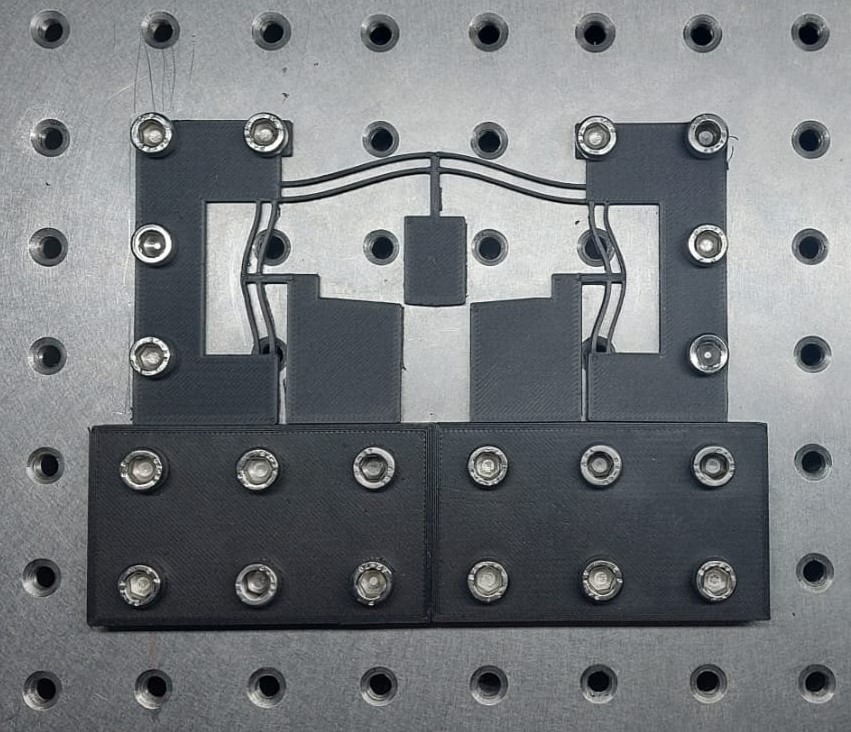} 
\caption{Initial position}
\label{fig_mechinitial}
\end{subfigure}
\begin{subfigure}{0.32\textwidth}
\includegraphics[width=\textwidth]{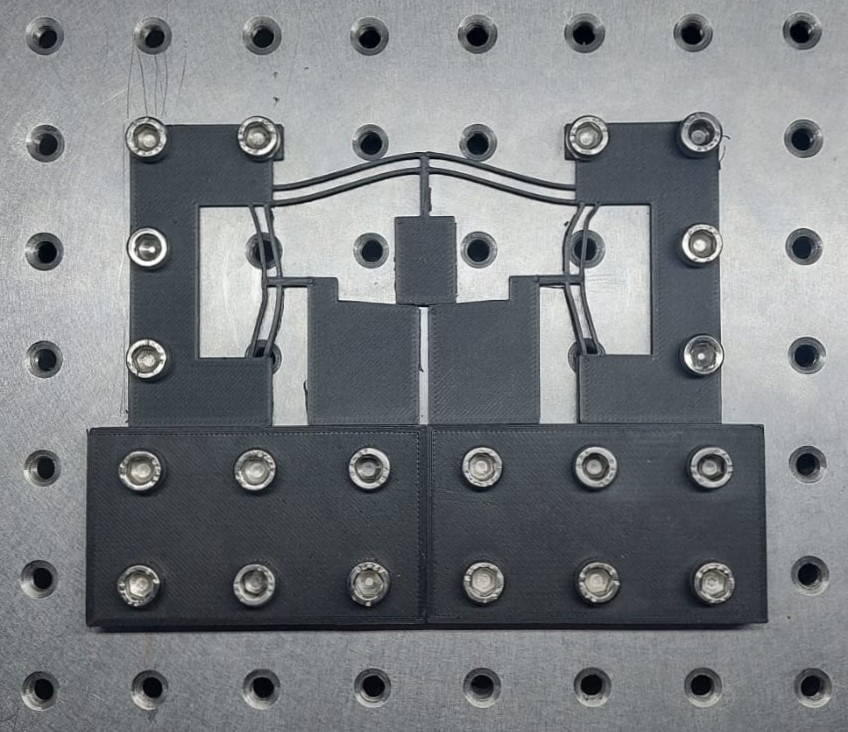}
\caption{Unlocked position}
\label{fig_mechunlocked}
\end{subfigure}
\begin{subfigure}{0.32\textwidth}
\includegraphics[width=\textwidth]{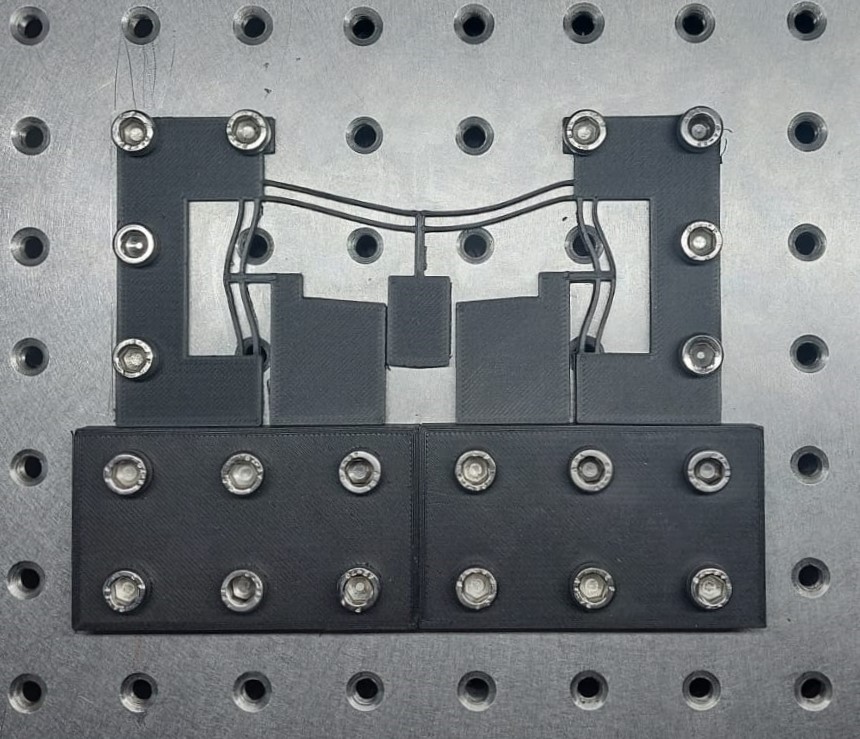}
\caption{Locked position }
\label{fig_mechlocked}
\end{subfigure}
\caption{Prototype of the model}
\label{fig_fullfinalmech}
\end{figure}

%Images of the prototype are given below. We 3D printed these models using Markforged 2 printer using Onyx material. Figure \Cref{fig:10.1} shows the mechanism in free state i.e., deformation is retained. Retained deformation is shown in \Cref{fig:10.2}. Figure \ref{fig:10.3} shows the mechanism in rigid state. 
A prototype was 3D-printed using Markforged 2 printer and Onyx material. \Cref{fig_mechinitial} shows the mechanism in its free state i.e., neither locked or unlocked. The unlocked state is shown in \cref{fig_mechunlocked} and \cref{fig_mechlocked} shows the mechanism in its locked state. 

\section{Summary}\label{sec6}

\begin{figure}[!htbp] 
    \centering
    \includegraphics[width=0.5\linewidth]{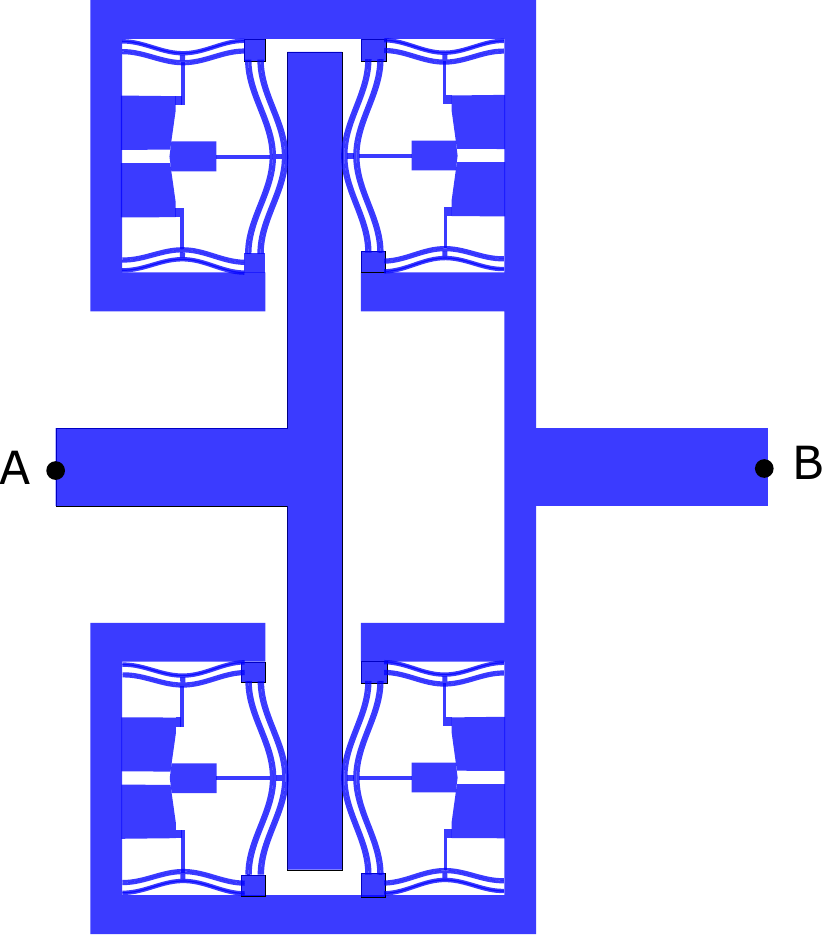}
    \caption{EDCM with four EDCMSs based on two bistable side arches and a perpendicular bistable central arch. }
    \label{fig_finalEDCM}
\end{figure}

The design of an EDCM by using bistable arches is presented, which could have potential applications to invert the Poisson's ratio of a compliant mechanism used as a biomechanical assay \cite{Manu2022} for cellular studies, to isolate a system from the surroundings from shocks and vibrations \cite{Xu2016}, and even meta-material fabrics with EDCM unit cells, which would cause the fabric to contract or expand laterally as the cloth is stretched.

The EDCMS is introduced as the critical design element of the EDCM and the four designs were presented. It was highlighted that the final mechanism is superior to the other three in terms of rigidity in the locked state and ease of switching between locked and unlocked states. This design involves two bistable arches on the side placed perpendicular to the horizontal central arch. EDCM made by using this EDCMS is illustrated in \cref{fig_finalEDCM}.

The design of the central and side arches were illustrated for their critical design parameters, such as the travel and the switching force. The arch dimensions were calculated using the formulae based on \cite{Palathingal2017}. The design was verified with the aid of FEA, 3D-printed, and demonstrated a working prototype.

%As mentioned before, the EDCM can reconfigure a given mechanism. It can be used as a potential bistable switch that could invert the Poisson's ratio of a compliant mechanism used as a biomechanical assay \cite{Manu2022} for cellular studies. It can also find applications to isolate a system from the surroundings from shocks and vibrations \cite{Xu2016}, and meta-material fabrics with EDCM in their unit cell, which would cause the fabric to contract or expand laterally as the cloth is stretched based on the engagement condition of the EDCM. 

\vspace{6 mm}
\bibliographystyle{plain}
\bibliography{sn-bibliography}

\vskip2pc

%

%\bibliography{refs}

\end{document}